\documentclass[11pt,a4paper]{amsart}
\usepackage{amsmath}
\usepackage{amssymb}
\usepackage{amsfonts}
\usepackage{hyperref}

\def\={\ =\ }

\def\e{{\,\rm e}\,}

\newcommand{\cn}{\mathcal{N}}

\theoremstyle{plain}

\numberwithin{equation}{section}
\newcommand{\beqa}{\begin{eqnarray}}
\newcommand{\eeqa}{\end{eqnarray}}
\newcommand{\beq}{\begin{equation}}
\newcommand{\eeq}{\end{equation}}

\begin{document}
\title[Supersymmetric gauge theories and Coulomb gases]{Supersymmetric gauge
theories, Coulomb gases \\
[5pt] and Chern-Simons matrix models}
\date{October 2013 \hfill \ EMPG--13--16 \ }
\urladdr{}
\author{Georgios Giasemidis}
\address{\flushleft Rudolf Peierls Centre for Theoretical Physics\\
University of Oxford\\
1 Keble Road, Oxford OX1 3NP, UK}
\email{g.giasemidis1@physics.ox.ac.uk}
\author{Richard J. Szabo}
\address{\flushleft Department of Mathematics\\
Heriot-Watt University\\
Colin Maclaurin Building, Riccarton, Edinburgh EH14 4AS, UK\\
Maxwell Institute for Mathematical Sciences, Edinburgh, UK\\
The Tait Institute, Edinburgh, UK}
\email{R.J.Szabo@hw.ac.uk}
\author{Miguel Tierz}
\address{\flushleft Departamento de An\'{a}lisis Matem\'{a}tico, Facultad de
Ciencias Matem\'{a}ticas \\
Universidad Complutense de Madrid \\
Plaza de Ciencias 3, Ciudad Universitaria, 28040 Madrid, Spain}
\email{tierz@mat.ucm.es}
\urladdr{}
\curraddr{ }
\subjclass{}
\keywords{}

\begin{abstract}
We develop Coulomb gas pictures of strong and weak coupling regimes of
supersymmetric Yang-Mills theory in five and four dimensions. By relating
them to the matrix models that arise in Chern-Simons theory, we compute
their free energies in the large $N$ limit and establish relationships
between the respective gauge theories. We use these correspondences to
rederive the $N^3$ behaviour of the perturbative free energy of
supersymmetric gauge theory on certain toric Sasaki-Einstein five-manifolds,
and the one-loop thermal free energy of $\mathcal{N}=4$ supersymmetric
Yang-Mills theory on a spatial three-sphere.
\end{abstract}

\maketitle

\section{Introduction and summary}

Recent interest in six-dimensional $(2,0)$ superconformal theories \cite%
{Witten} has been rekindled by the suggestion that maximally supersymmetric
Yang-Mills theory in five dimensions contains all degrees of freedom of the $%
(2,0)$ theory \cite{Douglas:2010iu, Lambert:2010iw,Bolognesi:2011nh}. The $%
(2,0)$ theory lives on the boundary of $AdS_{7}$, which in the Lorentzian
case can be chosen to be $S^{5}\times 
\mathbb{R}
$. The Euclidean version can have the time direction $%
\mathbb{R}
$ compactified to a circle $S^{1}$ which reduces the dual $(2,0)$ theory to
five-dimensional supersymmetric Yang-Mills theory. In \cite{Douglas:2010iu,
Lambert:2010iw,Bolognesi:2011nh} it is argued that the Kaluza-Klein states
from dimensional reduction over $S^{1}$ are mapped to instantons of the
five-dimensional gauge theory.

A well-known difficulty of the six-dimensional $(2,0)$ superconformal
theories is the lack of a Lagrangian description, and hence one has to use
the AdS/CFT correspondence where the $(2,0)$ theories are conjectured to be
dual to M-theory on an $AdS_{7}\times S^{4}$ background. This supergravity
dual is known to yield an $N^{3}$ growth in degrees of freedom for the free
energy of the $(2,0)$ theories~\cite{Klebanov:1996un,Henningson:1998gx}.
This dependence survives in the supergravity dual after compactification
suggesting that the $N^{3}$ behaviour should also appear in some way in the
five-dimensional gauge theory.

In \cite{Kim:2012ava} the $N^{3}$ behaviour is found by localization, which
reduces the partition function to one that is very close to the partition
function of Chern-Simons theory on $S^{3}$. In~\cite{Kallen:2012zn} the
calculation of the $\mathcal{N}=1$ supersymmetric Yang-Mills partition
function on $S^{5}$ is examined and, for the field theory with one adjoint
hypermultiplet which in the large radius limit has an enhanced $\mathcal{N}%
=2 $ supersymmetry, it is shown that the free energy scales as $N^{3}$
confirming the expectation from supergravity. The strong coupling limit is
studied in \cite{Kallen:2012zn} through the corresponding limit in the
matrix model description, which was found by localization in \cite%
{Kallen:2012va} (based on \cite{Kallen:2012cs}).

The partition function for the $\mathcal{N}=1$ supersymmetric gauge theory
on $S^5$ with gauge group $U(N)$ and a massless hypermultiplet in the
adjoint representation has the matrix model representation%
\begin{eqnarray}
Z_{\mathrm{YM}}^{(5)} &=& \int_{\mathbb{R}^N} \ \prod_{i=1}^{N}\, \mathrm{d}%
\phi _{i} \ {\mathrm{e}}\,^{-8\pi ^{3}\, r\, \phi _{i}^{2}/g_{\mathrm{YM}%
}^{2}}  \notag \\
&& \qquad \qquad \qquad \times \ \prod_{i<j}\, \left( \sinh \pi\, \phi
_{ij}\right) ^{2}\, \left( \cosh \pi \, \phi _{ij}\right) ^{1/2} \, \frac{%
\mathcal{S}_{3}({\,\mathrm{i}\,}\phi _{ij})}{\big(\mathcal{S}_{3}( \frac{1}{2%
}+{\,\mathrm{i}\,}\phi _{ij})\, \mathcal{S}_{3}( \frac{1}{2}-{\,\mathrm{i}\,}%
\phi _{ij}) \big)^{1/2} }  \label{N1-MM}
\end{eqnarray}%
where $r$ is the radius of the five-sphere $S^{5}$, $\phi _{ij}=\phi
_{i}-\phi _{j}$ with $\phi _{i}$ dimensionless matrix eigenvalues, and $%
\mathcal{S}_{3}(x)$ is the triple sine function which solves the equation~%
\cite{Kurokawa:1991} 
\begin{equation}
\frac{\mathrm{d} \log\mathcal{S}_{3}(x)}{\mathrm{d} x}=\pi \, x^{2}\, \cot
(\pi\, x) \ .
\end{equation}%
The matrix model represents the contribution to the localization formula for
the path integral around the trivial connection and hence gives the full
perturbative partition function, whereas the instanton sector contributes
with overall factors of order $\mathcal{O}\left( \exp (-16\pi ^{3}\, r/g_{%
\mathrm{YM}}^{2})\right) $.

In this paper we shall relate the strong coupling limit of (\ref{N1-MM}) to
the strong coupling expansion of the matrix model for $U(N)$ Chern-Simons
gauge theory on $S^{3}$, whose partition function is given by~\cite%
{Marino:2002fk} 
\begin{equation}
Z_{\mathrm{CS}} =\frac{{\,\mathrm{e}}\,^{-g_{\mathrm{s}}\, N\, (N^{2}-1)/12}%
}{N!}\, \int_{\mathbb{R}^N} \ \prod_{i=1}^{N}\, \frac{\mathrm{d} u_{i}}{2\pi 
} \ {\mathrm{e}}\,^{-u_{i}^{2}/2g_{\mathrm{s}}} \ \prod_{i<j}\, \Big(\,
2\sinh \frac{u_{i}-u_{j}}{2}\, \Big) ^{2} \ ,  \label{Zcs}
\end{equation}%
where $g_{\mathrm{s}}$ is the string coupling constant which is related to
the level $k\in\mathbb{Z}$ of the Chern-Simons gauge theory by $g_{\mathrm{s}%
}=2\pi {\,\mathrm{i}\,}/(k+N)$; in the following we work in the analytical
continuation of Chern-Simons theory with $g_{\mathrm{s}}$ real, as is done
in topological string theory~\cite{Gopakumar:1998vy}, which is the $q$%
-deformation of Yang-Mills theory on $S^2$~\cite{Szabo:2013vva}. This matrix
model also represents the contribution of the trivial flat connection, which
for Chern-Simons gauge theory on $S^{3}$ constitutes the complete
contribution to the path integral. Through this relation, we shall show how
to extract the $N^3$ dependence of the free energy directly from the
Chern-Simons matrix model using somewhat elementary techniques. This
relationship has the virtue of natually explaining certain aspects of
the exact localization of the five-dimensional supersymmetric gauge
theory; for example, we show that the contributions from adjoint
hypermultiplets to the localization formula in five dimensions can be
interpreted geometrically as a framing contribution of the three-manifold in the Chern-Simons partition function.

A key interpretation that we advocate from this relationship between
the two
apparently distinct gauge theories is through their natural appearences in
the theory of one-dimensional exactly solvable models. The partition
function (\ref{Zcs}) of Chern-Simons theory can be interpreted as the $\mathrm{L}^{2}
$-norm of the ground state wavefunction of a fermionic model on a
cylinder of radius $R_c=1$ with
Hamiltonian~\cite{Tierz:2008vh} 
\begin{equation}
H=-\sum_{i=1}^{N}\,\frac{\partial ^{2}}{\partial x_{i}^{2}}+\frac{1}{g_{%
\mathrm{s}}^{2}}\,\sum_{i=1}^{N}\,x_{i}^{2}+\frac{1}{g_{\mathrm{s}}\,R_{c}}%
\,\sum_{i<j}\,(x_{i}-x_{j})\,\coth
\Big(\,\frac{x_{i}-x_{j}}{2R_{c}}\,\Big) \ .
\label{H}
\end{equation}%
The strong coupling limit that identifies the two gauge theories is then a
thin cylinder limit $R_{c}\rightarrow 0$, wherein $\coth \big(\frac{x_{i}-x_{j}}{%
2R_{c}}\big)\rightarrow \mathrm{sgn}(x_{i}-x_{j})$, which as we shall see also identifies the
five-dimensional supersymmetric gauge theory with a one-dimensional
nonrelativistic charged Bose gas. With the aid of
some known Coulomb gas techniques, we are able to provide yet another
derivation of the $N^{3}$ behaviour of the free energy through relatively
straightforward methods. Our considerations of five-dimensional
supersymmetric Yang-Mills theory are contained in \S \ref{N=1SYM}.

The wavefunction of the fermionic model that appears in Chern-Simons theory
is the dimensional reduction of the Laughlin wavefunction on the cylinder,
which in the quantum Hall effect is the ground state of an electron gas in two
dimensions; in particular, the fermionic model lives on a longitudinal line
on the surface of the cylinder. Via this observation we use Coulomb gas
techniques to evaluate the one-loop thermal free energy of $\mathcal{N}=4$
supersymmetric Yang-Mills theory on $S^3\times S^1$, and reproduce the large 
$N$ results of~\cite{Hartnoll:2006pj} which interprets the gauge
theory effective action as a two-dimensional Coulomb gas in an
external potential; via holographic duality this finite-temperature gauge
theory can be used to relate weakly-coupled plasmas to black holes
and to map out stringy effects on the nature of black hole physics. A thin cylinder limit brings the
two-dimensional and one-dimensional models together, and it has been argued
that the two systems are adiabatically connected. Alternatively, by
constraining electrons in a strong magnetic field to the lowest Landau
level, the dimensional reduction can be achieved by taking a transversal
section of the cylinder as the space variable; via Fourier transformation, the
dual reduction is along a longitudinal line leading to the wavefunction of
the fermionic model. Hence the Chern-Simons matrix model also gives a
one-dimensional description of the quantum Hall effect on a cylinder. Our
considerations of four-dimensional supersymmetric Yang-Mills theory are the
topic of \S \ref{N=4SYM}.

\section{$\mathcal{N}=1$ supersymmetric Yang-Mills theory on $S^{5}$\label%
{N=1SYM}}

\subsection{Strong coupling regime}

In this section we shall focus on the strong coupling limit $%
\lambda\to\infty $ of supersymmetric Yang-Mills theory on $S^5$, where $%
\lambda := g_{\mathrm{YM}}^{2}\, N/r$ is the 't~Hooft coupling constant. In
this regime the partition function \eqref{N1-MM} takes the form~\cite%
{Kallen:2012zn} \footnote{%
Throughout we denote partition functions in their various limits, e.g.
strong and weak coupling limits, thermodynamic limits, etc., with a hat.} 
\begin{equation}
\hat Z_{\mathrm{YM}}^{(5)} = \int_{\mathbb{R}^N} \ \prod_{i=1}^{N}\, 
\mathrm{d}\phi _{i} \ \exp\Big({-\frac{8\pi ^{3}\, N}{\lambda } \,
\sum_{i=1}^{N}\, \phi _{i}^{2}+\frac{9\pi }{4}\, \sum_{i<j}\, |\phi
_{i}-\phi _{j}|} \Big) \ .  \label{Z_SYM}
\end{equation}%
The partition function \eqref{Z_SYM} is studied in \cite{Kallen:2012zn}
using the saddle-point method, exhibiting the $N^{3}$ behaviour of the free
energy at large $N$.

On the other hand, by rescaling the variables $u_{i}\rightarrow \sqrt{2g_{%
\mathrm{s}}} \, u_{i}$ the matrix integral \eqref{Zcs} in the strong
coupling limit $g_{\mathrm{s}}\rightarrow \infty $ becomes 
\begin{eqnarray}
&& \hat{Z}_{\mathrm{CS}} =\frac{{\,\mathrm{e}}\,^{-g_{\mathrm{s}}\, N\,
(N^{2}-1)/12}}{N!}\, \Big( \, \frac{g_{\mathrm{s}}}{2\pi ^{2}}\,\Big) ^{N/2}
\, \int_{\mathbb{R}^N } \ \prod_{i=1}^{N}\, \mathrm{d} u_{i} \ \exp\Big(%
-\sum_{i=1}^{N}\, u_{i}^{2}+\sqrt{2g_{\mathrm{s}}}\, \sum_{i<j}\,
|u_{i}-u_{j}| \Big) \ .  \label{Z_CS_shifted}
\end{eqnarray}
Hence the strong coupling limit of the $U(N)$ Chern-Simons matrix model on $%
S^{3}$ also reduces to (\ref{Z_SYM}). Since the original matrix model (\ref%
{N1-MM}) can be regarded as the Chern-Simons matrix model with additional
terms in the integrand and both matrix models have the same strong coupling
limit, the additional terms do not contribute in the strong coupling regime.
Thus the $N^{3}$ behaviour of the free energy should manifest itself in the
exact solution of the Chern-Simons matrix model. The $N^{3}$ dependence can
indeed be seen already in the solution of the Chern-Simons matrix model with
the technique of orthogonal polynomials \cite{Tierz:2002jj}; we shall show
below that the exact solution of the matrix model given in \cite%
{Tierz:2002jj} contains the exact evaluation of the matrix integral (\ref%
{Z_SYM}).

The computation of (\ref{Z_SYM}) is also intimately related to an old
statistical mechanics problem studied in detail by Baxter in 1963 \cite%
{Baxter:1963}. While the matrix model of $U(N)$ Chern-Simons gauge theory on 
$S^{3}$ has an interpretation as a one-component Coulomb plasma living on
the surface of a cylinder in Dyson's Coulomb gas picture of random matrix
ensembles~\cite{Tierz:2008vh}, the expression (\ref{Z_SYM}) is the partition
function of a one-dimensional Coulomb gas known as a one-dimensional jellium
model~\cite{Baxter:1963}: The two-body interaction $|x_i-x_j|$ is a Coulomb
potential in one dimension, while $\log\sinh(|x_i-x_j|/2R_c)$ is the Coulomb
potential on a cylinder of radius $R_c$. The strong coupling limit that maps
(\ref{Zcs}) to (\ref{Z_SYM}) is then a thin cylinder limit $R_c\to0$ in the
Coulomb plasma representation.

We can then compute \eqref{Z_SYM} in two different ways. First, we apply the
methods of Baxter in \cite{Baxter:1963} where the one-dimensional Coulomb
system with a uniform background charge distribution was studied. The second
method uses the exact solution of Chern-Simons gauge theory by examining its
strong coupling limit; in the process we will specify the framing
contribution contained in the matrix integral (\ref{Zcs}) which, in the
strong coupling limit, is the leading contribution.

\subsection{One-dimensional Coulomb gas}

\label{One-dimensional Coulomb gas}

The partition function \eqref{Z_SYM} is essentially a jellium model in one
dimension, studied in \cite{Baxter:1963}. This is a system of $N$
particles at temperature $T$ each carrying a charge $-\sigma $ on a line of length $2L$ with
particle density $\rho =(N-1)/2L$ and a uniform positive charge distribution 
$\rho \,\sigma $ along the line. The partition function of this system is
given by~\cite{Baxter:1963} 
\begin{equation}
Z_{\text{J}}^{(1)} = {\,\mathrm{e}}\,^{-N\, (N^{2}-1) /12\rho }\,
\int_{[-L,L]^N} \ \prod_{i=1}^{N}\, \mathrm{d} x_{i} \ \exp \bigg( -\frac{%
2\pi \, \sigma ^{2}}{T}\, \Big( \rho \, \sum_{i=1}^{N}\,
x_{i}^{2}-\sum_{i<j}\, |x_{i}-x_{j}|\Big) \bigg) \ ,  \label{ZJellium}
\end{equation}%
where the first and second terms in the exponential of the integrand
correspond to the charge-carrier--background interaction and charge carrier self-interactions,
respectively, while the proportionality term independent of $x_{i}$ is the
background--background interaction (i.e. the ground state energy).

Notice that \eqref{ZJellium} is directly related to \eqref{Z_SYM} under the
identifications $L=9\lambda/64\pi^2$ and 
\begin{equation}
\frac{\sigma^2}T= \frac98 \ , \qquad \rho =\frac{32\pi ^{2}\, N}{9\lambda} \
.  \label{identification_1}
\end{equation}
Therefore one can compute \eqref{Z_SYM} following the analysis of %
\eqref{ZJellium} in \cite{Baxter:1963}. We start by replacing the integral %
\eqref{Z_SYM} with the integral over the chamber of eigenvalue space with $%
\phi _{1}>\phi _{2}>\cdots >\phi _{N}$ to get 
\begin{equation}
\hat Z_{\mathrm{YM}}^{(5)}=N!\, \int_{-\infty }^{\infty }\, \mathrm{d}\phi
_{1} \ \int_{-\infty }^{\phi _{1}}\, \mathrm{d}\phi _{2} \ \cdots\
\int_{-\infty }^{\phi _{N-1}}\, \mathrm{d}\phi _{N} \ \exp \Big( -\frac{9\pi 
}{4}\, \sum_{i=1}^{N}\, \left( \rho \, \phi _{i}^{2}-\left( N-2i+1\right) \,
\phi _{i}\right) \Big) \ .
\end{equation}%
By completing the square in the potential term and setting $v_{i}=\rho \,
\phi _{i}+(2i-N-1)/2$ we find 
\begin{equation}
\hat Z_{\mathrm{YM}}^{(5)}={\,\mathrm{e}}\,^{\frac{9\pi }{48\rho }\,
N\,(N^2-1)}\, \frac{N!}{\rho ^{N}} \, \int_{-\infty }^{\infty }\, \mathrm{d}
v_{1} \ \int_{-\infty }^{v_{1}+1}\, \mathrm{d} v_{2} \ \cdots \
\int_{-\infty }^{v_{N-1}+1}\, \mathrm{d} v_{N}\ \exp\Big(-\frac{9\pi }{4\rho 
}\, \sum_{i=1}^{N}\, v_{i}^{2}\Big) \ .  \label{Z_SYM_baxter_0}
\end{equation}%
The multiple integral \eqref{Z_SYM_baxter_0} can be computed by first using
the approximation $v_{i}+1\simeq v_{i}$ for any $i$ at large $N$, and
employing the formula 
\begin{equation}
\int_{-\infty }^{v_{i-1}} \,\mathrm{d} v_{i} \ {\,\mathrm{e}}\,^{-c\,
v_{i}^{2}}\, \left( 1+\text{erf}(v_{i}\, \sqrt{c}\, )\right) ^{N-i}=\frac{%
\sqrt{\pi }}{2(N-i+1)\, \sqrt{c}}\, \left( 1+\text{erf}(v_{i-1}\, \sqrt{c}\,
)\right) ^{N-i+1}  \label{integral}
\end{equation}%
for $2\leq i\leq N$, where $\text{erf}(x) := \frac{2}{\sqrt{\pi }}\,
\int_{0}^{x}\, \mathrm{d} y \ {\,\mathrm{e}}\,^{-y^{2}}$ is the error
function and $c =9\pi /4\rho$ is a constant. The final integration over $%
v_{1}$ yields 
\begin{equation}
\int_{-\infty }^{\infty }\, \mathrm{d} v_{1} \ {\,\mathrm{e}}\,^{-c\,
v_{1}^{2}}\, \left( 1+\text{erf}(v_{1}\, \sqrt{c}\, )\right) ^{N-1}=\frac{%
2^{N-1}\, \sqrt{\pi }}{N\, \sqrt{c}} \ ,
\end{equation}%
and whence the partition function \eqref{Z_SYM_baxter_0} takes the form 
\begin{equation}
\hat Z_{\mathrm{YM}}^{(5)} = \frac{{\,\mathrm{e}}\,^{\frac{9\pi }{48\rho }%
\, N\,(N^{2}-1)}}{\rho ^{N}}\, N!\, \Big( \, \sqrt{\frac{{\pi }}{{c}}}\ %
\Big) ^{N-1}\, \Big(\, \prod_{i=2}^{N}\, \frac{1}{2(N-i+1)}\, \Big) \, \frac{%
2^{N-1}\sqrt{\pi }}{N\sqrt{c}}=\, \mathrm{e}\,^{\frac{9\pi }{48\rho }\,
N\,(N^{2}-1)}\, \Big(\, \frac{4}{9\rho }\, \Big) ^{N/2}  \label{Z_SYM_baxter}
\end{equation}
for $c=9\pi /4\rho $. The free energy of the strong coupling regime of the
supersymmetric gauge theory on $S^{5}$ at large $N$ is therefore given by 
\begin{equation}
\hat F_{\mathrm{YM}}^{(5)}=- \log \hat Z_{\mathrm{YM}}^{(5)}=-\frac{27}{512%
}\, \frac{g_{\mathrm{YM}}^{2}}{\pi\, r}\, N^{3} \ ,  \label{F_SYM_1}
\end{equation}%
in agreement with the result of~\cite{Kallen:2012zn}.

\subsection{Chern-Simons matrix model}

\label{Chern-Simons matrix model}

The matrix integral in (\ref{Z_CS_shifted}) is of the form %
\eqref{Z_SYM}: By rescaling the eigenvalues $\phi _{i}\rightarrow \sqrt{%
\lambda/{8\pi^3\, N}}\, \phi _{i}$ in \eqref{Z_SYM} and identifying the
large couplings as 
\begin{equation}
g_{\mathrm{s}} = \frac{81\lambda }{256\pi\, N}  \label{identification}
\end{equation}%
we have explicitly 
\begin{equation}
\hat Z_{\text{YM}}^{(5)} = \Big(\, \frac{8}{9}\, \Big) ^{N}\, N!\, \, {\,%
\mathrm{e}}\,^{g_{\mathrm{s}}\, N\, (N^{2}-1)/12} \ \hat{Z}_{\mathrm{CS}} \ .
\label{Z_SYM_2}
\end{equation}

The matrix model \eqref{Zcs} was solved in \cite{Tierz:2002jj} via the
Stieltjes-Wigert orthogonal polynomials, and its exact solution takes the
form 
\begin{equation}
Z_{\mathrm{CS}}=\Big(\, \frac{g_{\mathrm{s}}}{2\pi }\, \Big) ^{N/2}\, {\,%
\mathrm{e}}\,^{g_{\mathrm{s}}\, N\, (N^{2}-1)/12}\, \prod_{j=1}^{N}\, \big(%
1-q^{j} \big)^{N-j}  \label{CSexact}
\end{equation}%
where $q={\,\mathrm{e}}\,^{-g_{\mathrm{s}}}$. Applying the identification %
\eqref{identification} we now write \eqref{Z_SYM_2} as%
\begin{equation}
\hat Z_{\text{YM}}^{(5)}= N!\, \Big(\, \frac{\lambda }{8\pi ^{2}\, N}\, %
\Big) ^{N/2}\, {\,\mathrm{e}}\,^{\frac{27\lambda }{512\pi}\, (N^{2}-1)} \
\lim_{q\rightarrow 0}\ \prod_{j=1}^{N}\, \big(1-q^{j}\big)^{N-j} \ .
\label{limit}
\end{equation}%
The limit $q\to 0$ in (\ref{limit}), called the crystal limit in quantum
group theory~\cite{Kashiwara}, can be easily computed following~\cite%
{Forrester:1989vr}. For this, we take $\tilde\lambda=g_{\mathrm{s}}\, N$
constant and consider 
\begin{equation}
P(N,\tilde\lambda):=N\, \sum_{j=1}^{N}\, \Big( 1-\frac{j}{N}\Big) \, \log %
\Big( 1-\,\mathrm{e}\,^{-\tilde\lambda\, \frac{j}{N}}\Big) \ .
\end{equation}%
This expression is a Riemann sum over $y_{j}=\frac{j}{N}$
with $y_{j}-y_{j-1}=\frac{1}{N}$ for $j=1,\dots,N$. Since $\frac1N\leq
y_{j}\leq 1-\frac1N$, in the large $N$ limit we can write it as the integral 
\begin{equation}
P(N,\tilde\lambda)= N^{2}\, \int_{0}^{1}\, \mathrm{d} y \ \left( 1-y\right)
\, \log \big( 1-\,\mathrm{e}\,^{-\tilde\lambda \, y}\big) \ ,
\end{equation}%
and an additional change of variables $x=\tilde\lambda\, y$ in the limit $%
\tilde\lambda\to\infty$ gives finally 
\begin{equation}
\hat P(N,\tilde\lambda)= \frac{N^{2}}{\tilde\lambda }\, \int_{0}^{\infty }\, 
\mathrm{d} x \ \log \left( 1-\,\mathrm{e}\, ^{-x}\right) =-\frac{128 \pi
^{3}\,N^{2}}{243\lambda } \ .  \label{Sq=0}
\end{equation}%
It follows that 
\begin{equation}
\hat Z_{\text{YM}}^{(5)}=\Big( \, \frac{g_{\mathrm{YM}}^{2}}{4\pi ^{2}\, r}%
\, \Big) ^{N/2}\, N!\, \exp \Big(\, \frac{27}{512}\, \frac{g_{\mathrm{YM}%
}^{2}}{\pi \, r}\, N\,\big(N^{2}-1 \big)-\frac{128}{243}\, \frac{\pi ^{3}\, r%
}{g_{\mathrm{YM}}^{2}}\, N\, \Big)  \label{Z_SYM_3}
\end{equation}%
and the free energy at leading order in $N$ is given by 
\begin{equation}
\hat F_{\text{YM}}^{(5)}=-\frac{27}{512}\, \frac{g_{\mathrm{YM}}^{2}}{\pi\,
r} \, N^{3} \ ,  \label{F_SYM_2}
\end{equation}%
which coincides with \eqref{F_SYM_1}.

\subsection{Framing\label{Framing}}

The non-trivial part of the supersymmetric gauge theory partition function
on $S^5$, given by the product term in the Chern-Simons partition function (%
\ref{CSexact}), is subleading in $N$ and does not appear in the final result
of (\ref{F_SYM_2}). This naturally leads us into a discussion of the framing
contribution in Chern-Simons theory and how it is represented by the matrix
models.

Chern-Simons gauge theory is a theory of \textit{framed} knots and links 
\cite{Atiyah}. For gauge group $G=U(N)$, the contribution of a framing $%
\Pi_s $ on the three-sphere $S^3$ is parametrised by an integer $s\in 
\mathbb{Z}$ and takes the form \cite{Jeffrey:1992tk} 
\begin{equation}
\delta ( \Pi_s ) ={\,\mathrm{e}}\,^{2\pi{\,\mathrm{i}\,} s\, c/24}
\label{framing}
\end{equation}%
where $c=k\, \text{dim}(G)/(k+N)$ is the central charge of the WZW conformal
field theory based on the {affine extension} of $G$. The central charge can
be expressed in terms of the Weyl vector $\rho $ of the gauge group and one
has~\cite{Jeffrey:1992tk} 
\begin{equation}
\delta(\Pi_s ) ={\,\mathrm{e}}\,^{{\pi {\,\mathrm{i}\,} s\, |\rho |^{2}\, k}/%
{N\, (k+N)}}={\,\mathrm{e}}\,^{{\pi {\,\mathrm{i}\,} s\, |\rho |^{2}}/{N}}\, 
{\,\mathrm{e}}\,^{-{g_{\mathrm{s}}\, s\, |\rho |^{2}}/2} \ ,
\label{framing_2}
\end{equation}%
where we used the identification $g_{\mathrm{s}}=2\pi {\,\mathrm{i}\,}/(k+N)$
and 
\begin{eqnarray}
|\rho|^2 = \mbox{$\frac1{24}$}\, N\, \big(N^2-1\big) \ .
\end{eqnarray}
The inclusion of framing modifies the Chern-Simons partition function by
rescaling it with the phase (\ref{framing}), and one can therefore consider
a family of partition functions $Z_{\mathrm{CS}}^s$ parametrized by $s\in%
\mathbb{Z}$ with 
\begin{eqnarray}
Z_{\mathrm{CS}}^s = \delta(\Pi_s) \ Z_{\mathrm{CS}}^0 \ ,
\end{eqnarray}
where the partition function of Chern-Simons theory in the canonical framing 
$s=0$ on $S^3$ is given by%
\begin{equation}
Z_{\text{CS}}^{0}=\Big(\, \frac{g_{\mathrm{s}}}{2\pi } \, \Big) ^{N/2}\,
N!\, \prod_{j=1}^{N}\, \left( 1-q^{j}\right) ^{N-j} \ .  \label{Z_cf}
\end{equation}%
Thus the partition function \eqref{Zcs} carries a non-trivial framing
dependence, as is evident by comparing (\ref{Z_cf}) with (\ref{CSexact});
precisely, the Hermitian matrix model formulation of Chern-Simons gauge
theory on $S^3$ carries a framing contribution (\ref{framing}) with $s=-4$
such that 
\begin{equation}
Z_{\text{CS}}= \,\mathrm{e}\,^{-\pi {\,\mathrm{i}\,}(N^{2}-1)/6} \ Z_{\text{%
CS}}^{s=-4} \ .
\end{equation}

Let us consider now the strong coupling limit $g_{\mathrm{s}}\rightarrow
\infty .$ In this regime the five-dimensional supersymmetric gauge theory on
the boundary is related to the $q\rightarrow 0$ limit of the analytically
continued Chern-Simons theory with some framing contribution, through the
respective matrix integral formulations. Notice that in this strong coupling
limit, the framing dependence in Chern-Simons theory alters the prefactor of
the leading term in $N$, which is dominant in the limit. Therefore the $%
N^{3} $ behaviour of the $q\rightarrow 0$ limit of Chern-Simons theory comes
from the framing term, as the non-trivial product factor is subleading.

In fact, we can show that there exists an appropriate framing which explains
the discrepancy between the gauge theory and gravity results. The Yang-Mills
free energy from the matrix model computation is given by \eqref{F_SYM_2},
while the classical supergravity action in the $AdS_7$ background receives
contributions from the bulk, the boundary, and regularisation counterterms,
and is given by~\cite{Kallen:2012zn} 
\begin{equation}
\hat F_{\mathrm{grav}}= - \frac{5\pi\, R_6}{12 r}\, N^3
\end{equation}%
where $R_6$ is the radius of the compactification circle $S^1$ on the boundary.
The
Kaluza-Klein modes from compactification on $S^1$ are mapped to instantons of
the five-dimensional gauge theory, which suggests the identification \cite%
{Douglas:2010iu, Lambert:2010iw,Bolognesi:2011nh} 
\begin{equation}  \label{R6_vs_gym}
R_6 = \frac{g_{\mathrm{YM}}^{2}}{8 \pi^2}
\end{equation}
leading to 
\begin{equation}  \label{F_grav}
\hat F_{\mathrm{grav}}= -\frac5{96}\, \frac{g_{\mathrm{YM}}^{2}}{\pi\, r} \,
N^{3} \ .
\end{equation}
The mismatch in the numerical prefactors is restored by multiplying the
partition function \eqref{Z_SYM_3} of the boundary supersymmetric gauge
theory with an extra factor to give the partition function of a boundary
theory (that we denote by $\hat Z_{\mathrm{B}}$) which has the form 
\begin{eqnarray}
\hat Z_{\mathrm{B}} ={\,\mathrm{e}}\,^{-\frac{1}{1536}\frac{g_{\mathrm{YM}%
}^{2}}{\pi\, r} \, N^{3}} \ \hat Z_{\mathrm{YM}}^{(5)} = \Big(\, \frac{8}{9}%
\, \Big) ^{N}\, N!\ \hat Z_{\text{CS}}^{s=-8}\ {\,\mathrm{e}}\,^{\mathcal{O}%
(N^{2})} \ ,
\end{eqnarray}%
and the required framing parameter is $s=-640/81\simeq -8$. Alternatively,
since the leading term in the $q\to0$ limit comes from the second
exponential of the framing in \eqref{framing_2}, we equate 
\begin{equation}
\exp\Big(\, \frac{5g_{\mathrm{YM}}^{2}}{96\pi \, r}\, N^{3} \, \Big) =\exp%
\Big( \,-\frac{g_{\mathrm{s}}\, s\, N\,\big(N^{2}-1\big)}{48}\, \Big) \ ,
\end{equation}%
and by taking into account the identification (\ref{identification}) we find
that the boundary partition function can be expressed as the strong coupling
limit of the framed Chern-Simons partition function $Z^s_{\mathrm{CS}}$ with
framing parameter 
\begin{equation}
s=-\frac{5\cdot 256\cdot 48}{96\cdot 81}=-\frac{640}{81}\simeq -8 \ .
\end{equation}

\subsection{Massive hypermultiplet}

It was shown in \cite{Kallen:2012zn} that the $N^3$ behaviour in the
gauge theory on $S^5$ originates
from the presence of a single massless adjoint hypermultiplet, where
the field theory
has only $\mathcal{N}=1$ supersymmetry. In the case of a massive adjoint
hypermulitplet, it was argued in \cite{Kim:2012ava} that the global symmetry
is enhanced at a point where the mass is $M=\frac1{2r}$. Then the massless case
can be thought of as a deformation of the flat space theory by the
radius parameter $r$,
which in the large radius limit has an enhanced $\mathcal{N}=2$
supersymmetry. The massive case is considered in~\cite{Minahan:2013jwa},
where it was shown that the mass parameter enters into the numerical
prefactor of the free energy \eqref{F_SYM_2}. In particular, the strong
coupling limit of the partition function becomes 
\begin{equation}
\hat Z_{\mathrm{YM}}^{(5)}(m)= \int_{\mathbb{R}^N} \ \prod_{i=1}^{N}\, 
\mathrm{d}\phi _{i} \ \exp\bigg({-\frac{8\pi ^{3}\, N}{\lambda } \,
\sum_{i=1}^{N}\, \phi _{i}^{2}+\pi \, \Big(\, \frac{9}{4}+ m^2 \, \Big)\,
\sum_{i<j}\, |\phi _{i}-\phi _{j}|} \bigg) \ ,  \label{Z_SYM_m}
\end{equation}
where $m=-{\,\mathrm{i}\,} M\, r$ is the mass rotated to the imaginary axis, a
step required for the localization of the path integral. Hence the free
energy \eqref{F_SYM_2} is modified to 
\begin{equation}
\hat F_{\text{YM}}^{(5)}=-\Big(\, \frac{9}{4} + m^2\, \Big)^2\, \frac{g_{%
\mathrm{YM}}^{2}}{96 \pi\, r} \, N^{3} \ .  \label{F_SYM_2_m}
\end{equation}

The are now two key observations~\cite{Minahan:2013jwa}. First, the
matching of the supersymmetric Wilson loop that wraps the five-sphere $S^5$ at strong
coupling with the regularised circular Wilson loop in supergravity suggests
the new identification 
\begin{equation}  \label{new_identification}
R_6 = \frac{5 g_{\mathrm{YM}}^{2}}{32 \pi ^2} \ ,
\end{equation}
in contrast to \eqref{R6_vs_gym} which led to \eqref{F_grav}.
Second, one has to rotate back to real values of the mass parameter,
so that $%
m=\frac12$ at the enhancement point. This results in agreement between the free energy of five-dimensional
supersymmetric Yang-Mills theory and of its supergravity dual.

By suitably modifying the identifications of parameters in \eqref{identification_1} and \eqref{identification}, one
can easily obtain \eqref{F_SYM_2_m} via both the one-dimensional Coulomb gas
picture of \S\ref{One-dimensional Coulomb gas} and the strong coupling
regime of Chern-Simons theory from \S\ref{Chern-Simons matrix model}. In the
Chern-Simons description the $N^3$ behaviour originates from the
framing contribution whose choice controls the prefactor
containing the $N^3$ dependence, while from the point of view of the
supersymmetric gauge theory the $N^3$ dependence comes from the
presence of a single hypermultiplet whose mass parameter controls the prefactor of the free energy.
Applying the arguments of \S\ref{Framing} we should now equate 
\begin{equation}
\exp\Big(\, \frac{25 g_{\mathrm{YM}}^{2}}{384 \pi \, r}\, N^{3} \,
\Big) =\exp\Big( \,-\frac{g_{\mathrm{s}}\, s\, N\,\left(N^{2}-1\right)}{48%
}\, \Big) \ ,
\end{equation}
where the string coupling identification \eqref{identification} must be
modified to $%
g_{\rm s} = \frac{25 \lambda}{64 \pi\, N}$ in order to accommodate the dependence on the mass parameter $m=\frac12$. This yields the framing parameter $%
s=-8$. This is now an exact integer result, and it demonstrates the agreement of the strong
coupling regime of Chern-Simons theory with framing contribution $s=-8$ and
the large $N$ supergravity dual under the identification %
\eqref{new_identification}.
This consistency between Chern-Simons gauge theory on $S^3$ and 
supersymmetric Yang-Mills theory on $S^5$ raises the
intriguing possibility that there might be a deeper geometric connection between the
adjoint hypermultiplet of the five-dimensional gauge theory and
the framing contribution in Chern-Simons theory.

\subsection{Lens space matrix models}

For completeness, let us now generalise our computations to the strong
coupling regime $g_{\mathrm{s}}\to\infty$ of Chern-Simons theory on lens
spaces $L(P,Q)$, studied in~\cite{Marino:2002fk,Dolivet:2006ii}. The
contribution of the trivial flat connection to the path integral of this
gauge theory is described by the matrix model 
\begin{equation}  \label{Z_PQ}
Z_{\mathrm{CS}}^{P,Q} = \int_{\mathbb{R}^N}\ \prod_{i=1}^N\, \frac{\mathrm{d}
u_i}{2\pi} \ {\,\mathrm{e}}\,^{-u_i^2/2g_{\mathrm{s}}} \ \prod_{i<j} \left(
2\sinh \Big(\, \frac{u_i-u_j}{2P}\,\Big)\right)\, \left( 2\sinh \Big(\, 
\frac{u_i-u_j}{2Q}\, \Big)\right)
\end{equation}
where $P$ and $Q$ are coprime integers. For $P=Q=1$ this matrix integral is
related to the partition function (\ref{Zcs}) of $U(N)$ Chern-Simons theory
on $S^3$ as 
\begin{equation}  \label{Z_CS_S3}
Z_{\mathrm{CS}} = \frac{{\,\mathrm{e}}\,^{-g_{\mathrm{s}}\, N\, (N^2-1)/12}}{%
N!} \ Z_{\mathrm{CS}}^{1,1} \ .
\end{equation}
The matrix integral \eqref{Z_PQ} is computed exactly in \cite{Dolivet:2006ii}
via bi-orthogonal Stieltjes-Wigert polynomials with the result 
\begin{eqnarray}  
Z_{\mathrm{CS}}^{P,Q} &=& N! \, \Big(\frac{g_{\mathrm{s}}}{2\pi}\Big)^{N/2} \, 
\exp\Big(\mbox{$\frac{g_s}{2P^4}\, N \big[-\big(1+\frac{1}{2}\, (1+ \frac{P}{Q})\,
(N-1) \big)^2 +1+\frac{4}{3}\, \big(N^2-1\big) \big]$}\Big) \nonumber \\ &&
\times \ \prod_{j=1}^{N}\, \big(1-\bar
q\,^{j/P\, Q} \big)^{N-j}
\label{Z_PQ_sol}\end{eqnarray}
where $\bar q = {\,\mathrm{e}}\,^{-g_{\mathrm{s}}/P^2}$. We have already
seen that in the case of the three-sphere $P=Q=1$ the product contributes
subleading terms of order $N$ to the free energy in the limit $g_{\mathrm{s%
}}\to\infty$. The situation is the same for generic finite integers $%
P,Q$, and therefore the leading $N^3$ behaviour comes from the exponential
in the expression \eqref{Z_PQ_sol}. The corresponding free energy $\hat F_{%
\mathrm{CS}}^{P,Q}=-\log \hat Z_{\mathrm{CS}}^{P,Q} $ at large $N$ is given
by 
\begin{eqnarray}
\hat F_{\mathrm{CS}}^{P,Q} =- \frac{g_{\mathrm{s}}}{2P^4} \, \left ( \frac{13%
}{12}- \frac{P}{2Q}\, \Big(1+ \frac{P}{2Q} \, \Big) \right)\, N^3 \ .
\end{eqnarray}

Following the analogous manipulations for the $S^{3}$ matrix model, the
strong coupling limit of \eqref{Z_PQ} takes the form 
\begin{equation}
\hat{Z}_{\mathrm{CS}}^{P,Q}=\Big(\,\frac{g_{\mathrm{s}}}{2\pi ^{2}}\,\Big)%
^{N/2}\,\int_{\mathbb{R}^{N}}\ \prod_{i=1}^{N}\,\mathrm{d}u_{i}\ \exp \Big(%
-\sum_{i=1}^{N}\,u_{i}^{2}+\sqrt{\frac{g_{\mathrm{s}}}{2}}\,\alpha
\,\sum_{i<j}\,|u_{i}-u_{j}|\Big)  \label{Z_PQ_sc}
\end{equation}%
where $\alpha =\frac{1}{P}+\frac{1}{Q}$. It is tempting to compare this
partition function with that of supersymmetric Yang-Mills theory on the
squashed toric Sasaki-Einstein five-manifolds $Y^{P,Q}$~\cite{Gauntlett:2004yd} which was studied in~%
\cite{Qiu:2013pta}. In the limit of strong 't~Hooft coupling $\lambda =g_{%
\mathrm{YM}}^{2}\,N/r$, the equivariant perturbative partition function for
gauge group $U(N)$ and a massless matter hypermultiplet in the adjoint
representation simplifies to the matrix model 
\begin{equation}
\hat{Z}_{\mathrm{YM}}^{(5)}(P,Q)=\int_{\mathbb{R}^{N}}\ \prod_{i=1}^{N}\,%
\mathrm{d}\phi _{i}\ \exp \bigg(-\frac{8\pi ^{3}\,N\,\varrho }{\lambda
}\, \sum_{i=1}^N\, \phi_i^2 +%
\frac{\pi \,\varrho }{4}\,\Big(\,\mbox{$\sum\limits_{l=1}^4$}\,\omega _{l}\,%
\Big)^{2}\ \sum_{i<j}\,|\phi _{i}-\phi _{j}|\bigg)\ ,  \label{Z_SYM_shifted}
\end{equation}%
where $\varrho $ is the ratio of the equivariant volume of $Y^{P,Q}$ to the
volume of $S^{5}$ and $\omega _{1},\omega _{2},\omega _{3},\omega _{4}$ are
equivariant parameters for the isometric action of $U(1)^{4}$ on $\mathbb{C}%
^{4}$. Then (\ref{Z_SYM_shifted}) is proportional to (\ref{Z_PQ_sc}) under
the identification of the parameters 
\begin{equation}
\frac{\varrho }{128}\,\Big(\,\mbox{$\sum\limits_{l=1}^4$}\,\omega _{l}\,\Big)%
^{4}\,\frac{\lambda }{\pi \,N}=\frac{g_{\mathrm{s}}\,\alpha ^{2}}{2}\ .
\label{squashed-identification}
\end{equation}%
Using the identification (\ref{squashed-identification}) we can then write
the strong coupling free energy as 
\begin{equation}
\hat{F}_{\mathrm{CS}}^{P,Q}=-f(P,Q)\,\Big(\,\mbox{$\sum\limits_{l=1}^4$}%
\,\omega _{l}\,\Big)^{4}\,\varrho \,\frac{g_{\mathrm{YM}}^{2}}{\pi \,r}%
\,N^{3}\ ,
\end{equation}%
where 
\begin{equation}
f(P,Q)=\frac{1}{128\alpha ^{2}\,P^{4}}\,\left( \frac{13}{12}-\frac{P}{2Q}\,%
\Big(1+\frac{P}{2Q}\,\Big)\right) \ .
\end{equation}%
For $P=Q=1$ we get $f(1,1)=1/1536$, which agrees with the result of~\cite%
{Qiu:2013pta}; this corroborates the surprising universality of the $N^{3}$ behaviour
of the perturbative free energy on all five-manifolds $Y^{P,Q}$ that was
observed in~\cite{Qiu:2013pta}. The more general $L(P,Q)$ matrix models may
be related to a localization calculation of five-dimensional supersymmetric
Yang-Mills theory on the Sasaki-Einstein spaces $L^{a,b,c}$ which generalize 
$Y^{P,Q}$, but such a calculation is currently lacking in the literature and
is hence left for future work.

\subsection{One-dimensional wavefunctions\label{Wavefunction}}

The partition functions of some gauge theories on $S^{3}$
can be written as the norm or the overlap of some one-dimensional quantum
mechanical wavefunctions. This is true of the $\cn=4$ theories on
$S^3$ that arise as the low-energy limit of $\cn=4$ supersymmetric
Yang-Mills theory in four dimensions~\cite{Nishioka:2011dq}, which is
also connected to the six-dimensional $(2,0)$ superconformal theories
via certain dimensional reductions. It is also the case of
Chern-Simons gauge theory \cite%
{Tierz:2008vh}. 

For the one-dimensional wavefunction
\begin{equation}
\Psi _{0}( x_{1},\ldots,x_{N}) =\prod_{i=1}^{N}\, \e^{-\frac{%
\omega }{2}\, x_{i}^{2}} \ \prod_{i<j}\, \exp \Big(\, \frac{c\, \left\vert
x_{i}-x_{j}\right\vert }{2}\, \Big) \label{GaussianLL}
\end{equation}%
the same direct approach in \cite{Tierz:2008vh} can be used to find the
general Hamiltonian of a bosonic model for which (\ref{GaussianLL}) is
a ground state. It
can also be found as a limit of the Hamiltonian for the Chern-Simons fermionic
model (\ref{H}) which in its most general version is characterized by a
ground state wavefunction \cite{Tierz:2008vh}%
\begin{equation}
\Psi _{0}^{(m)}( x_{1,},\ldots,x_{N}) =\prod_{i=1}^{N}%
\, \e^{-{x_{i}^{2}}/{2g_{\rm s}}} \ \prod_{i<j}\, \Big(\, \sinh \frac{x_{i}-x_{j}}{%
2R_c}\, \Big) ^{m} \ ,  \label{fermionG}
\end{equation}%
where $m$ is a positive parameter. The corresponding
Hamiltonian is%
\begin{eqnarray}
H_m &=&-\sum_{i=i}^{N}\, \frac{\partial ^{2}}{\partial x_{i}^{2}}+\frac{1}{%
g_{\rm s}^{2}}\, \sum_{i=1}^{N}\, x_{i}^{2}+\frac{m}{g_{\rm s}\,
R_c}\, \sum_{i<j}\, (x_{i}-x_{j})%
\, \coth \Big(\, \frac{x_{i}-x_{j}}{2R_c}\, \Big)   \notag \\
&&+\, \frac{m\, (m-1)}{2R_c}\, \sum_{i<j}\, \frac{1}{\sinh ^{2}\big( \frac{x_{i}-x_{j}}{2R_c%
}\big) } \ .  \label{fermionH}
\end{eqnarray}%
For $m=1$ we obtain the Hamiltonian $H=H_1$ in (\ref{H}). 
The bosonic and fermionic models are related through the limit $R_c\rightarrow
0$, which as discussed earlier is a
thin cylinder limit. First, let us see what happens to the two-body term of the
wavefunction in the limit%
\begin{equation}
\lim_{R_c\to0}\, \Big(\, \sinh \frac{x_{i}-x_{j}}{2R_c}\, \Big) ^{m}=
2^{-m}\, \big( 
\mathrm{sgn}(x_{i}-x_{j})\big) ^{m}\, \exp \Big(\, \frac{m\, \left\vert
x_{i}-x_{j}\right\vert }{2R_c}\, \Big) \ .
\end{equation}%
If $m$ is even the sign term does not appear. Having $%
m$ odd and keeping the term $\mathrm{sgn}(x_{i}-x_{j})$ can be interpreted
as a fermionization of the resulting boson wavefunction, in the sense of 
\cite{Girar}. We can now identify $c=\frac m{R_c}$ with the usual parameter
of the Lieb-Liniger model~\cite{LL}. To obtain generic values of $c$
in the thin cylinder limit, we need to take $m\rightarrow 0$, in which
case the sign terms above disappear. Thus in the
limit $R_c\rightarrow 0$ the wavefunction of the fermionic model (\ref{fermionG})
reduces to (\ref{GaussianLL}) (up to normalization)
with $\omega=\frac1{g_{\rm s}}$ and $c=\frac m{R_c}$. The Hamiltonian
(\ref{fermionH}) correspondingly becomes%
\begin{equation}
H_0=-\sum_{i=i}^{N}\, \frac{\partial ^{2}}{\partial
  x_{i}^{2}}+\frac{1}{g_{\rm s}^{2}} \,
\sum_{i=1}^{N}\, x_{i}^{2}+\frac{c}{g_{s}}\, \sum_{i<j}\, \left\vert
x_{i}-x_{j}\right\vert + 4c\, (m-1)\, \sum_{i<j}\, \delta ( x_{i}-x_{j}) \ .
\label{H2}
\end{equation}%
The Hamiltonian (\ref{H2}) can be regarded as a
generalization of the Lieb-Liniger model~\cite{LL}, although the special
case that appears in Chern-Simons theory is (\ref{fermionG}) with $m=1$.
Therefore, in the limit considered above, it leads to the charged Bose gas
without delta-function interactions. For this model the Coulomb gas
interpretation holds for both the Hamiltonian and the Dyson Coulomb gas
picture of the wavefunction, since both cases involve the
one-dimensional Coulomb potential
$\left\vert x_{i}-x_{j}\right\vert $.

\section{$\mathcal{N}=4$ supersymmetric Yang-Mills theory on $S^{3}\times
S^{1}$\label{N=4SYM}}

\subsection{Weak coupling regime}

We shall focus now on $\mathcal{N}=4$ supersymmetric Yang-Mills theory on $%
S^{3}\times S^{1}$ and its representation as a Coulomb gas on $\mathbb{R}%
\times S^1$ at weak (but finite) 't~Hooft coupling $\lambda =g_{\text{YM}%
}^{2}\, N$; the radius of $S^3$ is denoted $R$ and the inverse radius of $%
S^1 $ is the temperature $T$. Based on the one-loop determinants computed in 
\cite{Hollowood:2006xb}, the effective action in the low temperature limit $%
R\, T\ll 1$ is shown by~\cite{Hartnoll:2006pj} to become 
\begin{equation}
\hat S_{\text{YM}}^{(4)}[z,\bar z] = N^{2}\, \Big(\, \frac{3\beta }{16R}%
-\log 2\Big) + \frac{N\, \pi ^{2}\, R}{\beta \, \lambda }\, \sum_{i=1}^{N}\, %
\big(z_{i}+\bar{z}_{i}\big)^{2}-2\sum_{i<j}\, \log \big|\sinh (z_{i}-z_{j}) %
\big|  \label{S_lowT}
\end{equation}%
where $\beta =\frac1T$ while $z_{i}=\frac12\, (\beta \, \phi _{i}+{\,\mathrm{%
i}\,}\theta _{i})$ are complex scalar fields with $\phi _{i}$ and $\theta
_{i}$ the eigenvalues of the adjoint scalar fields and of the temporal component
of the gauge field, respectively. The action \eqref{S_lowT} describes a
Coulomb gas on a cylinder of radius $R_{c}=\frac12$.

The Coulomb gas on the cylinder is intimately related to the Coulomb gas in
one dimension, as has been studied in detail in \cite%
{RezHald,Sei1,Sei2,Kar,BKWHK,Jansen:2007}, often in the context of the
corresponding quantum Hall effect on the cylinder. In particular, it is
found that for certain Hamiltonians a quantum Hall ground state does not
undergo a phase transition when the two-dimensional surface of the system is
deformed in a quasi-one-dimensional (thin cylinder) limit~\cite%
{Sei1,Sei2,Kar,BKWHK}. Hence the two-dimensional and one-dimensional systems
are argued to be adiabatically connected. Recall that the $U(N)$\
Chern-Simons matrix model on $S^{3}$ admits an interpretation as a Coulomb
system of restricted dimension, i.e. its interaction is the Coulomb
interaction on the cylinder but the particles live in one dimension (a
longitudinal line on the surface of the cylinder)~\cite{Tierz:2008vh}. In
fact, the charge-density wave behaviour of the Laughlin wavefunction on the
cylinder~\cite{RezHald} is also manifest in the oscillatory behaviour of the
density of states of the Chern-Simons matrix model~\cite{deHaro:2005rz}. The
same property is analyzed more rigorously in \cite{Jansen:2007} as an
example of translational symmetry breaking. Either by dimensional reduction
(which is achieved as usual by projecting to the lowest Landau level with a
limit of large magnetic field $B\rightarrow \infty $) or by a thin cylinder
limit, the two Coulomb gas descriptions are directly related. This suggests
a relationship between the respective gauge theories, even though the nature
of their Coulomb gas description is rather different. In the case of
Chern-Simons gauge theory the matrix model arises exactly due to a
localization of the path integral on flat connections \cite{Beasley:2005vf},
whereas in the case of $\mathcal{N}=4$ supersymmetric Yang-Mills theory it
follows from an effective field theory approach together with a number of
simplifications such as the description of the condensate of the scalar
fields in a single coordinate~\cite{Hartnoll:2006pj}.

\subsection{Two-dimensional Coulomb gas}

\label{2DCoulomb}

The corresponding partition function takes the form 
\begin{eqnarray}
\hat Z_{\text{YM}}^{(4)} &:= &\int_{(\mathbb{R}\times S^1)^N}\
\prod_{i=1}^{N} \, \mathrm{d}^{2}z_{i} \ {\,\mathrm{e}}\,^{-\hat S_{\text{YM%
}}^{{(4)}}[z,\bar z]}  \notag \\[4pt]
&=&\Big(\,\frac{2^{N+1}}\beta\,\Big)^N \, {\,\mathrm{e}}\,^{-3\beta\, N^{2}
/16R} \ \int_{\mathbb{R}^{N}} \ \prod_{i=1}^{N}\, \mathrm{d} x_{i} \
\int_{[0,\pi ]^{N}} \ \prod_{i=1}^{N}\, \mathrm{d} y_{i} \ {\,\mathrm{e}}%
\,^{-\tau\, x_i^2} \ \prod_{i<j}\, \big|\sinh (z_{i}-z_{j}) \big|^2
\label{Z1}
\end{eqnarray}%
where $z_i=x_i+{\,\mathrm{i}\,} y_i$ are coordinates on the cylinder and we
set $\tau := 4 \pi ^{2}\, N\, R/\beta\, \lambda$ for brevity. The product in
the integrand can be written as 
\begin{eqnarray}
\prod_{i<j}\, \big|\sinh (z_{i}-z_{j})\big|^{2} = 2^{-N\, (N-1)}\,
\prod_{i=1}^{N}\, {\,\mathrm{e}}\,^{-2(N-1)\, x_{i}} \ \prod_{j<k}\,
\left\vert {\,\mathrm{e}}\,^{2z_{j}}-{\,\mathrm{e}}\,^{2z_{k}}\right\vert
^{2} \ .  \label{elogsinh}
\end{eqnarray}%
We now complete the square in the exponential in $x_i$ and shift variables $%
x_{i}\rightarrow x_{i}- (N-1)/\tau $, which implies $z_{i}\rightarrow
z_{i}-(N-1)/\tau $, and then rescale $z_i\rightarrow z_i /\sqrt{\tau }$ so
that the partition function finally takes the form 
\begin{eqnarray}
\hat Z_{\text{YM}}^{(4)} &=&\Big(\, \frac{4}{\beta\,\tau}\, \Big)^N\, {\,%
\mathrm{e}}\,^{-3\beta\, N^{2} /16R}\, {\,\mathrm{e}}\,^{-N\, (N-1)^{2}/\tau
}  \notag  \label{Z_N_4} \\
&&\times \ \int_{\mathbb{R}^{N}} \ \prod_{i=1}^{N}\, \mathrm{d} x_{i} \
\int_{[0,\sqrt{\tau }\, \pi ]^{N}} \ \prod_{i=1}^{N}\, \mathrm{d} y_{i} \ {\,%
\mathrm{e}}\,^{-x_{i}^{2}} \ \prod_{i<j}\, \big\vert \, {\,\mathrm{e}}%
\,^{2z_{i}/\sqrt{\tau }}-{\,\mathrm{e}}\,^{2z_{j}/\sqrt{\tau }}\, \big\vert %
^{2} \ .
\end{eqnarray}

\subsection{Laughlin wavefunction}

It is possible to analyze the partition function in the same spirit as
\S\ref{Wavefunction} by using known
properties of the Laughlin wavefunction on the cylinder. The Laughlin
wavefunction is the ground state wavefunction of a two-dimensional electron
gas in a uniform neutralising background with a uniform magnetic field; it
was introduced to describe the fractional quantum Hall effect~\cite%
{Laughlin:1983}. The Laughlin wavefunction for the cylinder was first
considered in \cite{Thouless} but did not become an object of further study
until later on, beginning with~\cite{RezHald}; mathematical aspects,
such as
its translational symmetry breaking, were studied in~\cite{Jansen:2007}. It takes the
form 
\begin{equation}
\Psi _{N}(z;\gamma_B,p):=\frac{{\,\mathrm{e}}\,^{-p^{2}\, \gamma_B^{2}\, N\,
(N-1)\, (2N-1)/12}}{\sqrt{N!}}\, \Big(\, \frac{\gamma_B}{2\pi^{3/2}}\, \Big) %
^{N/2} \ \prod_{i=1}^N\,{\,\mathrm{e}}\,^{-x_{i}^{2}/2} \ \prod_{j<k}\, %
\big( {\,\mathrm{e}}\,^{\gamma_B \, z_{j}}-{\,\mathrm{e}}\,^{\gamma_B \,
z_{k}}\big) ^{p}  \label{Laughlin's function}
\end{equation}%
where $z_i=x_i+{\,\mathrm{i}\,} y_i$ represents the coordinates of the
fermions on the cylinder, $p$ is the filling fraction of the quantum Hall
system and $\gamma_B$ is a dimensionful parameter defined as the ratio of
the magnetic length $\ell_{B}=\left( \hbar /e\, B\right) ^{1/2}$ (here set
equal to $1$) to the radius of the cylinder (here $R_{c}=\frac12$). Its $%
\mathrm{L}^{2}$-norm is given by 
\begin{equation}
C_{N}(\gamma _{B},p):=\big\|\Psi _{N}\big\|_2^{2}=\int_{\mathbb{R}^{N}} \
\prod_{i=1}^N\, \mathrm{d} x_i \ \int_{[0,2\pi /\gamma _{B}]^{N}} \
\prod_{i=1}^N\, \mathrm{d} y_i \ \big\vert \Psi _{N}(z;\gamma _{B},p)%
\big\vert ^{2} \ .
\end{equation}

We now notice that the gauge theory partition function \eqref{Z_N_4} can be
expressed in terms of the normalisation constant $C_{N}(\gamma _{B},p)$ for $%
\gamma _{B}=\frac2{\sqrt{\tau }}$ and $p=1$ as 
\begin{eqnarray}
\hat Z_{\text{YM}}^{(4)} = \Big(\, \frac{4\pi^{3/2}}{\beta\, \sqrt{\tau }}%
\, \Big) ^{N} \, N!\, {\,\mathrm{e}}\,^{-3\beta\, N^{2} /16R}\, {\,\mathrm{e}%
}\,^{N\, (N^{2}-1)/3\tau}\ C_{N}\big(\gamma _{B}=%
\mbox{$\frac2{\sqrt{\tau
}}$}\,,\,p=1\big) \ .  \label{Zlt_CN}
\end{eqnarray}%
For $p=1$ the $\mathrm{L}^2$-norm is given by~\cite{Jansen:2007} \footnote{%
For $p=1$ the Laughlin wavefunction becomes a Slater determinant, which is
the wavefunction of $N$ fermions. Then $C_{N}=\|\Psi _{N}\|_2^{2}=1$ is the
normalisation of the wavefunction of $N$ electrons.}%
\begin{equation}
C_{N}(\gamma _{B},p=1)=1 \ .
\end{equation}%
By substituting back $\tau = 4\pi ^{2}\, N\, R/\beta \, \lambda$ the free
energy at large $N$ is thus given by%
\begin{equation}
\hat F_{\text{YM}}^{(4)} =-\log \hat Z_{\text{YM}}^{(4)} =\Big( \, \frac{3%
}{16}-\frac{\lambda }{12\pi ^{2}}\, \Big) \, \frac{N^{2}\, \beta }{R} \ ,
\label{free_energy}
\end{equation}%
in agreement with the calculation of the free energy given in \cite%
{Hartnoll:2006pj} in the Coulomb gas description.

\subsection{Jellium on the cylinder}

\label{2D jellium}

The partition function \eqref{Z_N_4} can also be computed exactly by mapping
the problem to a one-component plasma on the cylinder, known as the
two-dimensional jellium model, at the fermion coupling $\Gamma=2 \gamma $
which was studied in \cite{Choquard:1983} for $\gamma = 1$ and in \cite%
{Samaj:2004} for arbitrary integer values of $\gamma$. The two-dimensional
jellium model is defined as follows. Consider $N$ particles of charge $%
q$ on a cylinder of radius $R_c$ and finite length $L$ embedded in a
homogeneous background of charge density $\rho_b=-q\, n$, where $n = N/(2\pi\,
L\, R_c)$ so that the system remains neutral. The partition function takes
the form \cite{Samaj:2004} 
\begin{equation}
Z_{\text{J}}^{(2)} = \frac{1}{N!} \, \int _{\Lambda^N}\ \prod_{i=1}^N \, 
\mathrm{d}^2 z_i {\ \mathrm{e}}\,^{-\beta \, E_N[z,\bar z]} \ ,
\label{Z_2D jellium_stat}
\end{equation}
where $\Lambda = \big[-\frac L2, \frac L2 \big] \times [-\pi,\pi]$ is the
cylinder and the total energy of the system is given by 
\begin{equation}  \label{Energy_2Djellium}
E_N[z,\bar z] = \pi \, n\, q^2\, \sum_{i=1}^N\, x_i^2 - q^2 \, \sum _{i<j}
\, \log \Big |2\sinh \frac{z_i-z_j}{2R_c} \Big| + B_N \ .
\end{equation}
The first and second sums correspond to the charge-carrier--background and
charge-carrier--charge-carrier interactions, respectively, while the third
term $B_N$ which is independent of $z_i$ corresponds to the
background--background interaction. The fermion coupling is defined as the
dimensionless combination $\Gamma= \beta \, q^2$, and after some simple
algebra analogous to that of \S \ref{2DCoulomb} the partition function is
written as 
\begin{equation}  \label{Z_2D_jellium}
Z_{\text{J}}^{(2)} = \frac{1}{N!} \, \int _{\Lambda^N} \ \prod_{i=1}^N \, 
\mathrm{d}^2 z_i \ w (z_i,\bar z_i) \ \prod _{i<j} \, \big| {\,\mathrm{e}}%
\,^{z_i/R_c} - {\,\mathrm{e}}\,^{z_j/R_c}\big|^{\Gamma}
\end{equation}
where $w (z, \bar z)$ is the one-particle Boltzmann factor given by 
\begin{equation}  \label{w}
w(z, \bar z) = w (x) = \frac{1}{4\pi^2\,R_c^2} \, {\,\mathrm{e}}\, ^{-\pi \,
n \, \Gamma\, (x^2 + x\, \frac{N-1}{2\pi\, n \, R_c})} \ .
\end{equation}
Completing the square in the Boltzmann factor of \eqref{Z_2D_jellium},
shifting the variables $x_i\to x_i- \frac{N-1}{4\pi\, n \, R_c} $, and then
rescaling variables $z_i \to z_i/\sqrt{\pi\, n\, \Gamma}$ we finally get 
\begin{eqnarray}  \label{Z_2D_jellium_2}
Z_{\text{J}}^{(2)} &=& \frac{1}{N!} \, \frac{1}{\big(8\pi^3 \, n\, \gamma\,
R_c^2 \big)^N} \, {\,\mathrm{e}}\, ^{-\frac{\gamma \, \pi}{8\pi^2\, n \,
R_c^2}\, N\, (N-1)^2}  \notag \\
&& \times \ \int _{\Lambda^N} \ \prod _{i=1} ^{N} \, \mathrm{d}^2 z_i \ {\,%
\mathrm{e}}\, ^{- x_i^2} \ \prod _{i<j}\, \bigg | {\,\mathrm{e}}\,^{\sqrt{%
\frac{1}{2\pi\,n \, \gamma \, R_c^2}}\, z_i} - {\,\mathrm{e}}\,^{\sqrt{\frac{%
1}{2\pi\,n \, \gamma \, R_c^2}}\, z_j}\bigg |^{2 \gamma} \ ,
\end{eqnarray}
where we replaced $\Gamma = 2 \gamma$ for the general case following \cite%
{Samaj:2004}.

The gauge theory partition function \eqref{Z_N_4} is proportional to the
partition function of the two-dimensional jellium model %
\eqref{Z_2D_jellium_2} in the thermodynamic limit $N,L\rightarrow \infty $
with $n$ constant,\footnote{%
In this limit, the integration volume becomes 
\begin{eqnarray*}
\int_{\Lambda^N } \ \prod_{i=1}^{N}\,\mathrm{d}^{2}z_{i}=2^{N}\, \int_{%
\mathbb{R}^{N}}\ \prod_{i=1}^N\, \mathrm{d} x_i \ \int_{[0,\pi ]^{N}}\
\prod_{i=1}^N\, \mathrm{d} y_i
\end{eqnarray*}%
where we used the fact that the integrand in $\mathrm{Im}(z_i)=y_i$ is an
even function.} for $\gamma =1$ with the identification 
\begin{equation}
\tau = 8\pi\, n\, R_c^{2} \ ,  \label{identification_3}
\end{equation}
as 
\begin{equation}
\hat Z_{\text{YM}}^{(4)}={\,\mathrm{e}}\,^{-3\beta\, N^{2}/16R}\, \pi^{2N}
\ \hat Z_{\text{J}}^{(2)}(\gamma =1) \ .
\end{equation}%
The partition function \eqref{Z_2D_jellium_2} is computed for various values
of $\gamma $ in \cite{Samaj:2004}, and in particular for $\gamma =1$ it
takes the form 
\begin{equation}
Z_{\text{J}}^{(2)}(\gamma =1)=\prod_{j=0}^{N-1}\, \Big(\,\frac{1}{2\pi\,R_c}%
\, \int_{-L/2}^{L/2}\, \mathrm{d} x \ {\,\mathrm{e}}\,^{-2\pi \, n\,
x^{2}+(2j-(N-1))\, x/R_c} \, \Big) \ .
\end{equation}%
In the thermodynamic limit the integral is Gaussian and we find 
\begin{equation}
\hat Z_{\text{J}}^{(2)}(\gamma =1)=\Big( \, \frac{1}{8\pi^2\, R_c^{2}\, n}\, %
\Big) ^{N/2} \, {\,\mathrm{e}}\,^{\frac{1 }{24\pi\, R_c\, n}\, N\,
(N^{2}-1)} \ .  \label{largeN}
\end{equation}
Via \eqref{largeN} we can now compute the large $N$ limit of the partition
function \eqref{Z_N_4} for the low temperature limit of $\mathcal{N}=4$
supersymmetric Yang-Mills theory at finite weak coupling, and we find 
\begin{equation}
\hat Z_{\text{YM}}^{(4)} =\Big(\, \frac{\pi^{3/2}}{\sqrt{\tau }}\, \Big) %
^{N/2}\, N!\, {\,\mathrm{e}}\,^{-3\beta\,N^{2} /16R+N\, (N^{2}-1)/3\tau} \ .
\label{Z_SYM_6}
\end{equation}%
The partition function \eqref{Z_SYM_6} is identical, up to a proportionality
factor $2^{N}$, to the partition function \eqref{Zlt_CN}, and therefore the
free energy in the large $N$ limit reads 
\begin{equation}
\hat F_{\text{YM}}^{(4)} =\Big( \, \frac{3}{16}-\frac{\lambda }{12\pi ^{2}}%
\, \Big) \, \frac{N^{2}\, \beta }{R}+{\mathcal{O}}(N)
\end{equation}%
which coincides with \eqref{free_energy}.

In the Coulomb gas description of~\cite{Hartnoll:2006pj}, the low
temperature distribution of the eigenvalues lies uniformly in a band of
width $2A$ and circumference $\pi $. This is consistent with the
two-dimensional jellium picture on a cylinder of length $L$ and
circumference $2\pi\, R_c$: The identification \eqref{identification_3} can
be written as 
\begin{equation}
\frac{\beta \,\lambda }{2\pi ^{2}\, R}=\frac{L}{2R_c}=2A \ ,
\end{equation}%
in agreement with the result of~\cite{Hartnoll:2006pj}. This coincidence can
be substantiated by noticing that the interpretation of the gauge theory
effective action \eqref{S_lowT} as a Coulomb gas in an external potential
considered in~\cite{Hartnoll:2006pj} is in fact the two-dimensional jellium
model. This is already apparent in \eqref{S_lowT}, where the term $\log
\sinh |z_{i}-z_{j}|$ corresponds to the interaction potential between the
charge-carriers, the term $x_{i}^{2}$ is related to the charged
particle--background interaction, and the $z_{i}$-independent term of order $%
N^{2}$ is proportional to the background--background interaction constant $%
B_N $~\cite{Samaj:2004}. Whence the external potential in the Coulomb gas
picture is the background--background interaction in the two-dimensional
jellium description.

\subsection{Dimensional reduction}

We shall now study the dimensional reduction of the Laughlin wavefunction on
the cylinder and the thin cylinder limit of the two-dimensional jellium
system. In both cases we end up with the Coulomb gas description of the
Chern-Simons matrix model.

In \cite{Azuma:1993ra} one can find an explicit relationship between the Laughlin
state of the quantum Hall effect and certain one-dimensional exactly
solvable models with long-range interactions such as the Calogero model and
the Sutherland model. In the limit of a strong magnetic field $B\to\infty$,
the charge-carriers in two dimensions are constrained to the lowest Landau
level and two of the four phase space degrees of freedom freeze, reducing
the number of effective degrees of freedom to two, one in space representation and one
in momentum representation. Depending on the two-dimensional geometry of the
Hall system, the one-dimensional representation of the Laughlin ground state
corresponds to the ground state of either the Calogero model (for the disc)
or the Sutherland model (for the cylinder); in the latter case the axial
degrees of freedom freeze (in space representation) \cite{Azuma:1993ra}.
However, instead of the axial degrees of freedom one can also dually reduce
the periodic degrees of freedom by working in momentum representation. The
Laughlin ground state with filling factor $p$ on the cylinder is of the form %
\eqref{Laughlin's function} with the change of coordinates $z_{j}\rightarrow
z_{j}/\sqrt{B}$ and restoring the $B$-dependence on $\gamma _{B}=\frac{%
\ell_{B}}{R_{c}}=1/\sqrt{B}\, R_{c}$ in units where $\hbar =e=1$. Then the
one-dimensional reduction in momentum representation of the Laughlin state
on a cylinder is given by \cite{Azuma:1993ra} 
\begin{eqnarray}
\langle t_{1},\ldots, t_{N}|\Psi_N \rangle =\prod_{i=1}^N\, \exp\Big(-\frac{1%
}{2B}\, \frac{\partial^2 }{\partial t_{i}^2} \Big) {\,\mathrm{e}}\,^{-B\,
t_{i}^{2}/2} \ \prod_{j<k}\, \big({\,\mathrm{e}}\,^{t_{j}/R_{c}}-{\,\mathrm{e%
}}\,^{t_{k}/R_{c}} \big)^{p} \ ,  \label{1Drep_Laughlin}
\end{eqnarray}%
where $t_{i}$ is the eigenvalue of the eigenstate $|t_{i}\rangle $ of the
operator $X_{i}=x_{i}+\Pi _{y_i}/B$ for the guiding centre coordinate of the
cyclotron motion.\footnote{%
Our notation differs from that of \cite{Azuma:1993ra}, where $x_i$ denote
the periodic coordinates and $y_i$ the axial coordinates.} In the limit of
strong magnetic field $B\rightarrow \infty $, this wavefunction
reduces to 
\begin{equation}
\prod_{i=1}^N\, {\,\mathrm{e}}\,^{-B\, (t_{i}-t_{0})^{2}} \ \prod_{j<k}\,
\left( \sinh \Big( \frac{2(t_{j}-t_{k})}{R_{c}}\Big) \right) ^{p}
\end{equation}%
where $t_{0}=p\, (N-1)/2B\, R_{c}$. As in \S\ref{Wavefunction}, this is the wavefunction of a
one-dimensional model with interaction potential $\sinh ^{-2}(x_{i}-x_{j})$.
An intriguing consequence of the strong magnetic field limit is that the
wavefunction of the one-component plasma in one dimension with this
interaction potential for filling factor $p=1$ is related to the
Chern-Simons matrix model \eqref{Zcs}, as shown
by~\cite{Tierz:2008vh}.\footnote{An alternative perspective on the
relationship between Chern-Simons gauge theory on $S^3$ and the
Sutherland model can be found in~\cite{Szabo:2010qv}.}

Using the Laughlin wavefunction interpretation of the low temperature limit
of supersymmetric Yang-Mills theory in four dimensions, we can apply the
strong magnetic field limit to the expression \eqref{Zlt_CN}. This suggests
that we should identify the magnetic field $B$ with the quantity $\tau $ so
that 
\begin{equation}
B= \tau =\frac{4\pi ^{2}\, N\, R\, T}{\lambda } \ .  \label{B=tau}
\end{equation}%
Therefore the strong magnetic field limit corresponds to $\tau\gg 1$. We
should then take into account the domain of validity of the effective action %
\eqref{S_lowT} from~\cite{Hartnoll:2006pj}, which is determined at weak
't~Hooft coupling $\lambda$ via one-loop perturbation theory~\cite%
{Hollowood:2006xb}. There it was argued that the perturbative calculation is
valid for the range of temperatures with 
\begin{equation}
0\leq R\,T\ll \mbox{$\frac{1}{\lambda }$}
\end{equation}%
at weak coupling $\lambda $. This range is satisfactory for high
temperatures because the radius of the spatial sphere $S^{3}$ provides a
natural infrared cutoff of order $R\sim 1/\sqrt{\lambda }\, T$. However,
there is no restriction on $R\, T$ for low temperatures. From \eqref{B=tau}
it follows that the large $\tau $ limit is valid only for low temperatures
of order of $\lambda $, i.e. $R\, T\gtrsim \lambda $ at large $N$, and it
might break down for low temperatures of order $R\, T\ll \lambda $.

The geometrical meaning of the strong magnetic field limit can be deduced in
momentum representation where the axial degrees of freedom on the cylinder
are kept and the periodic ones are frozen \cite{Azuma:1993ra}. In this
description, the two boundaries of the cylindrical Laughlin droplet are
placed at $X_{1}=0$ and $X_{2}=p\, (N-1)/B\, R_{c}$. Taking $B\rightarrow
\infty $ requires sending $R_{c}\rightarrow 0$ so that $X_{2}$ is constant
and the cylinder does not collapse to a circle. Thus the strong magnetic
field limit freezes the radial degrees of freedom reducing the geometry of
the cylinder effectively to one dimension.

\section*{Acknowledgments}

We thank Sabine Jansen for correspondence on Coulomb gases.
The work of GG was partially supported by the A.G.~Leventis Foundation and
the A.S.~Onassis Public Benefit Foundation Grant F-ZG 097/ 2010-201. GG
thanks the Department of Mathematical Analysis, Universidad
Complutense de Madrid for the kind hospitality during his visit. The work of
RJS was partially supported by the Consolidated Grant ST/J000310/1 from the
UK Science and Technology Facilities Council, and by Grant RPG-404 from the
Leverhulme Trust. The work of MT has been partially funded by the
Project Entanglement in Quantum
Systems (MTM2011-26912), the Project QUITEMAD: QUantum Information
TEchnologies MADrid (S2009/ESP-1594) and a Juan de la Cierva Fellowship.

\end{document}